\documentclass[]{article}

\usepackage{savesym}
\usepackage{graphicx}
\usepackage{authblk}
\usepackage{lineno}
\usepackage[percent]{overpic}
\usepackage{xcolor}
\usepackage{hyperref}
\savesymbol{tablenum}
\usepackage{siunitx}

\restoresymbol{SIX}{tablenum}
\usepackage[version=3]{mhchem}
\usepackage{makecell}
\usepackage{geometry}
\usepackage[utf8]{inputenc}       
\usepackage[T1]{fontenc}         
\usepackage{amsmath}
\usepackage{amssymb}
\usepackage{mathrsfs}             
\usepackage{lineno}
\usepackage{tikz}
\usetikzlibrary{plotmarks,patterns,calc,decorations.markings}
\DeclareMathAlphabet{\mathpzc}{OT1}{pzc}{m}{it}
\usepackage{booktabs}
\usepackage{lipsum}
\usepackage[symbol]{footmisc}

\catcode`\@=11 
\def\parenbar{\mathpalette\p@renb@r}
\def\p@renb@r#1#2{\vbox{%
\ifx#1\scriptscriptstyle \dimen@.7em\dimen@ii.2em\else
\ifx#1\scriptstyle \dimen@.8em\dimen@ii.25em\else
\dimen@1em\dimen@ii.4em\fi\fi \offinterlineskip
\ialign{\hfill##\hfill\cr
\vbox{\hrule width\dimen@ii}\cr
\noalign{\vskip-.3ex}%
\hbox to\dimen@{$\mathchar300\hfil\mathchar301$}\cr
\noalign{\vskip-.3ex}%
$#1#2$\cr}}}
\catcode`\@=12 %
\def\nuan{\parenbar{\nu}\kern-0.4ex}

\newlength{\smfigwidth}
\newlength{\figwidth}
\newlength{\captwidth}
\usepackage{hyperref}
\usepackage[all]{hypcap}

\title{Search for neutrinos from the tidal disruption events AT2019dsg and AT2019fdr with the ANTARES telescope}

\date{}

\author[1,2]{A.~Albert}
\author[3]{S.~Alves}
\author[4]{M.~Andr\'e}
\author[5]{M.~Anghinolfi}
\author[6]{G.~Anton}
\author[7]{M.~Ardid}
\author[8]{J.-J.~Aubert}
\author[9]{J.~Aublin}
\author[9]{B.~Baret}
\author[10]{S.~Basa}
% ...

\affil[{ }]{(ANTARES Collaboration)}

\affil[1]{\scriptsize{Universit\'e de Strasbourg, CNRS,  IPHC UMR 7178, F-67000 Strasbourg, France}}
\affil[2]{\scriptsize{Universit\'e de Haute Alsace, F-68200 Mulhouse, France}}
\affil[3]{\scriptsize{IFIC - Instituto de F\'isica Corpuscular (CSIC - Universitat de Val\`encia) c/ Catedr\'atico Jos\'e Beltr\'an, 2 E-46980 Paterna, Valencia, Spain}}
\affil[4]{\scriptsize{Technical University of Catalonia, Laboratory of Applied Bioacoustics, Rambla Exposici\'o, 08800 Vilanova i la Geltr\'u, Barcelona, Spain}}
\affil[5]{\scriptsize{INFN - Sezione di Genova, Via Dodecaneso 33, 16146 Genova, Italy}}
\affil[6]{\scriptsize{Friedrich-Alexander-Universit\"at Erlangen-N\"urnberg, Erlangen Centre for Astroparticle Physics, Erwin-Rommel-Str. 1, 91058 Erlangen, Germany}}
\affil[7]{\scriptsize{Institut d'Investigaci\'o per a la Gesti\'o Integrada de les Zones Costaneres (IGIC) - Universitat Polit\`ecnica de Val\`encia. C/  Paranimf 1, 46730 Gandia, Spain}}
\affil[8]{\scriptsize{Aix Marseille Univ, CNRS/IN2P3, CPPM, Marseille, France}}
\affil[9]{\scriptsize{Universit\'e de Paris, CNRS, Astroparticule et Cosmologie, F-75006 Paris, France}}
\affil[10]{\scriptsize{Aix Marseille Univ, CNRS, CNES, LAM, Marseille, France }}
% ...

\begin{document}

\maketitle

\begin{abstract}
On October~1,~2019, the IceCube Collaboration detected a muon track neutrino with high probability of being of astrophysical origin, IC191001A. After a few hours, the tidal disruption event (TDE) AT2019dsg, observed by the Zwicky Transient Facility (ZTF), was indicated as the most likely counterpart of the IceCube track. More recently, the follow-up campaign of the IceCube alerts by ZTF suggested a second TDE, AT2019fdr, as a promising counterpart of another IceCube muon track candidate, IC200530A, detected on May~30,~2020. These are the second and third associations between astrophysical sources and high-energy neutrinos after the compelling identification of the blazar TXS~0506+056. Here, the search for ANTARES neutrinos from the directions of AT2019dsg and AT2019fdr using a time-integrated approach is presented. As no significant evidence for space clustering is found in the ANTARES data, upper limits on the one-flavour neutrino flux and fluence are set. 
\end{abstract}

\textit{Introduction - } High-energy cosmic neutrinos are a unique probe of the high-energy universe: their detection is critical to identify the acceleration sites of cosmic rays and to discover distant sources otherwise inaccessible with $\gamma$-rays. Several milestones have been achieved in the field of neutrino astronomy in the past decade, which also opened new unresolved questions.
\newline \indent The IceCube Collaboration has detected an isotropic high-energy cosmic neutrino flux with a high level of significance in several diffuse flux searches~\cite{ICHESE3years, ICMuon6years}. A mild excess, consistent with the IceCube findings, has been seen also in the ANTARES data~\cite{ANTDIFF9y}. However, the sources responsible for the diffuse cosmic signal have not been yet identified.
\newline \indent In 2017, a $\sim$200~TeV IceCube muon neutrino was found positionally coincident with the direction of a known blazar in a flaring state, TXS 0506+056~\cite{TXSAlert, ICmultimessenger}. Soon after, the IceCube Collaboration reported a compelling evidence for an earlier neutrino flare from the same direction found in the archival data~\cite{ICTXS}. This led to the first association of high-energy neutrinos with an astrophysical source. 
However, the expected neutrino emission from the blazar TXS~0506+056 can only contribute to the IceCube cosmic neutrino flux for less than 1\%~\cite{ICTXS}. 
The region around TXS~0506+056 was studied also by the ANTARES Collaboration, yielding no significant correlation in time and/or space with the IceCube neutrinos~\cite{ANTARESTXS}.
\newline \indent In 2020, a new likely association between an IceCube neutrino, IC191001A, and an astrophysical source, the tidal disruption event (TDE) AT2019dsg, was announced~\cite{TDEassociation}. Soon after, a second TDE, ZTF19aatubsj/AT2019fdr, hereafter referred to as AT2019fdr, was found positionally coincident with another IceCube neutrino, IC200530A~\cite{ZTFfdr}.
TDEs are rare transient phenomena that occur when a star passes close enough to a supermassive black hole and is disrupted by the strong tidal forces (for recent reviews see~\cite{TDErev1, TDErev2}).
The accretion process onto the black hole involves about half of the stellar mass and, in some cases, results in a relativistic jet. Consequently, both jetted and non-jetted TDEs are observed.
Since TDEs show non-thermal radio emission, they  have been suggested as promising sources of ultra-high energy cosmic rays and high-energy neutrinos (see~\cite{TDEneutrinoREV} and references therein).
Nonetheless, the contribution to the IceCube diffuse flux from the TDEs detected before AT2019dsg and AT2019fdr has been constrained to be at most $\sim$1.3$\%$ ($\sim$26$\%$) in the jetted (non-jetted) TDE case~\cite{SteinProc}. 
\newline \indent The detection of IC191001A, a $\sim$200~TeV muon neutrino with $59\%$ probability of being of astrophysical origin, registered by IceCube on October~1,~2019~\cite{GoldBronzeEvents, IceCubeAlert}, triggered dedicated follow-ups by multiple telescopes, including the Zwicky Transient Facility (ZTF)~\cite{Zwicky, Zwicky2}. While several optical transients were found within the $90\%$~localisation uncertainty of the neutrino direction ($\sim$26~$\mathrm{deg}^{2}$) by ZTF, the TDE AT2019dsg, being one of the few ever detected radio-emitting TDEs, was soon identified as the most promising candidate neutrino source~\cite{IceCubeAlert, TDEassociation}. 
Analogously, IC200530A, a $\sim$80~TeV muon neutrino with $59\%$ probability of being of astrophysical origin, detected by IceCube on May~30,~2020~\cite{GoldBronzeEvents, IceCubeAlert200530A}, was followed by ZTF, which indicated AT2019fdr as the likely counterpart of the neutrino event~\cite{ZTFfdr}.
\newline \indent AT2019dsg and AT2019fdr were first discovered by ZTF on April~9,~2019~\cite{TDEdiscovery} and on April~27, 2019~\cite{AT2019fdrSpectrum}, respectively. The TDEs showed a luminosity peak in the optical/UV bands, and appeared to have reached a luminosity plateau by the time of the neutrino detections, which happened $\sim$150~days and about~one~year after the peak, respectively~\cite{TDEassociation, AT2019fdrSpectrum}. As detailed in~\cite{TDEassociation}, after the discovery, AT2019dsg was also detected in several follow-up observations in optical/UV, X-rays and radio, indicating particle acceleration to relativistic energies. The probability of finding by chance a radio-emitting TDE as bright as AT2019dsg in bolometric energy flux and in coincidence with the IceCube neutrino was estimated to be $0.2\%$. The expected neutrino flux from AT2019dsg has been calculated for different models of multi-messenger emission both in the jetted~\cite{TDEWinter, TDELiu} and non-jetted~\cite{TDEMurase} TDE cases. The most optimistic scenarios predict a neutrino fluence of $\sim$10$^{-2}$~GeV~cm$^{-2}$. 
\newline \indent In this letter, the ANTARES follow-up of these findings is presented. In particular, the ANTARES data collected since the discovery of AT2019dsg and AT2019fdr are investigated to look for a steady neutrino emission from the direction of the TDEs.

\textit{Detector and data sample - } The ANTARES neutrino telescope~\cite{antaresdetector} is a three-dimensional array of 885~photomultiplier tubes (PMTs) located 40~km off-shore from Toulon, France, below the surface of the Mediterranean Sea. The 10-inch~PMTs are distributed along 12, 450~m long, vertical lines anchored to the sea-bed at a depth of about 2500~m and held taut by a buoy at the top, instrumenting a total volume of $\sim$0.01$\ \textrm{km}^3$. The neutrino detection principle is based on the collection of the Cherenkov light induced by relativistic charged particles produced in neutrino interactions within and around the instrumented volume. The information provided by the position, time and collected charge of the signals in the PMTs is used to infer the direction and energy of the incident neutrino. 
\newline \indent Different neutrino flavours and interactions leave distinct signatures in the detector. Two event topologies can be identified in the ANTARES telescope: tracks and showers. Tracks are originated by the passage in water of relativistic muons produced in charged current (CC) interactions of muon neutrinos. The long lever arm of the track topology allows to reconstruct the parent neutrino direction of high-quality selected events with a median angular resolution of~$0.4^{\circ}$ for energies above $100$~TeV~\cite{latestPS}.
All-flavour neutral current (NC) as well as ${\nu_e}$ and ${\nu_\tau}$~CC interactions induce showers, characterised by an almost spherical light emission around the shower maximum, with an elongation of a few metres. A median angular resolution of~$\sim$3$^{\circ}$ for high-quality selected events with energies between 1~TeV and 0.5~PeV is achieved for this topology~\cite{latestPS}.
\newline \indent The search for neutrinos from the direction of AT2019dsg (AT2019fdr) presented in this letter includes both track-like and shower-like events recorded in ANTARES from the day of the discovery of the TDE, 2019 April 9 (2019 April 27), until the day of the last available fully-calibrated ANTARES data, 2020 February 29, corresponding to a livetime of 315 (298) days. 
The track and shower events are selected using the criteria described in~\cite{latestPS}. The selection is optimised to minimise the neutrino flux needed for a $5\sigma$ discovery of a point-like source emitting with an energy spectrum $\propto E_{\nu}^{-2.0}$. Cuts are applied on the zenith angle, the angular error estimate and parameters describing the quality of the reconstruction. In the shower channel, an additional cut is applied on the interaction vertex, required to be located within a fiducial volume slightly larger than the instrumented volume. A total of 413 (390) tracks and 8 (7) showers, resulting from the selection, are employed in the search for a cluster of events from the direction of AT2019dsg (AT2019fdr).

\textit{Search method - } The location of AT2019dsg ($\mathrm{RA} =314.26^{\circ}$, $\delta = 14.20^{\circ}$) and of AT2019fdr ($\mathrm{RA} =257.28^{\circ}$, $\delta = 26.86^{\circ}$) has been investigated to search for spatial clustering of events above the known background expectation following an unbinned maximum likelihood ratio approach. The likelihood describes the ANTARES data in terms of signal and background probability density functions (PDFs) and is defined as:

\begin{align} \label{eq:lik}
    \log \mathcal L = \sum_{\mathcal J\in{\{\it{tr}, \it{sh}\}}} \sum_{i\in\mathcal J} \log \Big[ \mu_\mathrm{sig}^{\mathcal J}
       {\mathcal S}^{\mathcal J}_{i} + {\mathcal N}^{\mathcal J}  {\mathcal B}^{\mathcal J}_{i}\Big]
    - \mu_\mathrm{sig} ,
\end{align} 

\noindent where ${\mathcal S}^{\mathcal J}_{i}$ and ${\mathcal B}^{\mathcal J}_{i}$ are the values of the signal and background PDFs for the event $i$ in the sample ${\mathcal J}$ ($\it{tr}$ for tracks, $\it{sh}$ for showers), $\mu_\mathrm{sig}^{\mathcal J}$ and ${\mathcal N}^{\mathcal J}$ are respectively the number of unknown signal events and the total number of data events in the $\mathcal J$ sample, and $\mu_\mathrm{sig} = \mu_\mathrm{sig}^{\it{tr}} + \mu_\mathrm{sig}^{\it{sh}}$ is the total number of fitted signal events. The PDFs have been built using the 2007-2020 ANTARES data and Monte Carlo production in order to ensure sufficient statistics. 
The signal and background PDFs are given by the product of a directional and an energy-dependent term. While atmospheric neutrino events appear randomly distributed on the sky, neutrinos from point-like sources are expected to accumulate around the source position. The energy information helps to distinguish signal from background, as a softer energy spectrum is predicted for atmospheric neutrinos with respect to the expected signal~\cite{ANTATMO}. An unbroken power law neutrino spectrum, $\Phi_{\nu}(E_{\nu}) \propto E_{\nu}^{-\gamma}$, is assumed in this analysis for the signal emission, with $\gamma$ being one of the three tested spectral indices, $\gamma = 2.0$, $\gamma = 2.5$, or $\gamma = 3.0$. The same definition of the PDFs used in the ANTARES 9-year point-source analysis~\cite{latestPS} is employed in this search. The spatial signal PDF is a spline parametrisation of the point-spread function. It is defined as the PDF to reconstruct an event at a given angular distance from its original direction due to reconstruction uncertainties, and it is derived from Monte Carlo simulations of $\propto E_{\nu}^{-\gamma}$ cosmic neutrinos. The background is assumed to be uniform in right ascension, and the distribution of the sinus of the declination of the selected data is employed as spatial background PDF. Monte Carlo simulations of $\propto E_{\nu}^{-\gamma}$ energy spectrum cosmic neutrinos (signal) assuming neutrino flavour equipartition at Earth and of atmospheric neutrinos using the spectrum of~\cite{atmonu} (background) are used to derive the energy PDFs. 
In the likelihood maximisation, the number of signal events, $\mu_\mathrm{sig}$, is the only free parameter. The test statistic, $\mathcal Q$, is defined from the likelihood as: 

\begin{equation}
    \mathcal Q = 2 \log \left[ \frac{\mathcal L(\mu_\mathrm{sig} = \hat{\mu}_\mathrm{sig})}{ \mathcal L(\mu_\mathrm{sig} = 0)}\right],
    \label{eq:teststat}
\end{equation}

\noindent where $\hat{\mu}_\mathrm{sig}$ is the best-fit value that maximises the likelihood.
The significance of the potential spatial cluster at the location of the TDE is estimated by means of pseudo-experiments (PEs) -- pseudo-data sets of data randomised in time to eliminate any local clustering due to potential sources. The fraction of background-only PEs with a value of $\mathcal Q$ larger than the one obtained with the data gives the significance (p-value) of the cluster.
\newline \indent The 5$\sigma$ discovery potential of the search, $n^{5\sigma}$, is defined as the mean number of injected signal events needed for a 5$\sigma$ discovery in 50\% of the PEs. The obtained values are $n^{5\sigma} = 3.6$ and $n^{5\sigma} = 3.4$ for AT2019dsg and AT2019fdr, respectively, assuming an $E_{\nu}^{-2.0}$ neutrino spectrum, while $\sim$40$\%$~more events are needed for the softer spectrum $E_{\nu}^{-3.0}$.

\begin{figure*}[ht]
\centering
\begin{overpic}[width=0.49\textwidth]{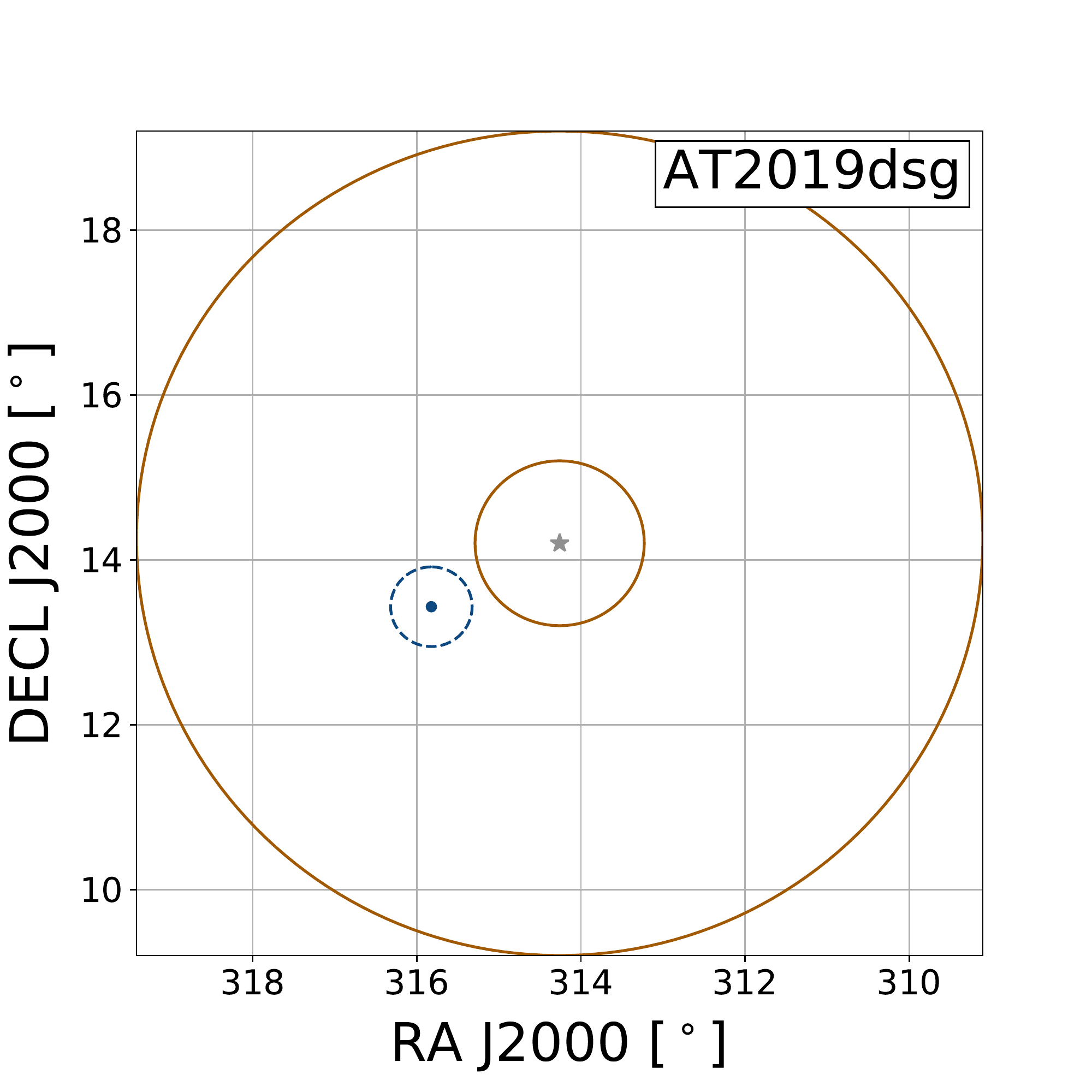}
\put (13,89){\color{black} PRELIMINARY}
\end{overpic}
\begin{overpic}[width=0.49\textwidth]{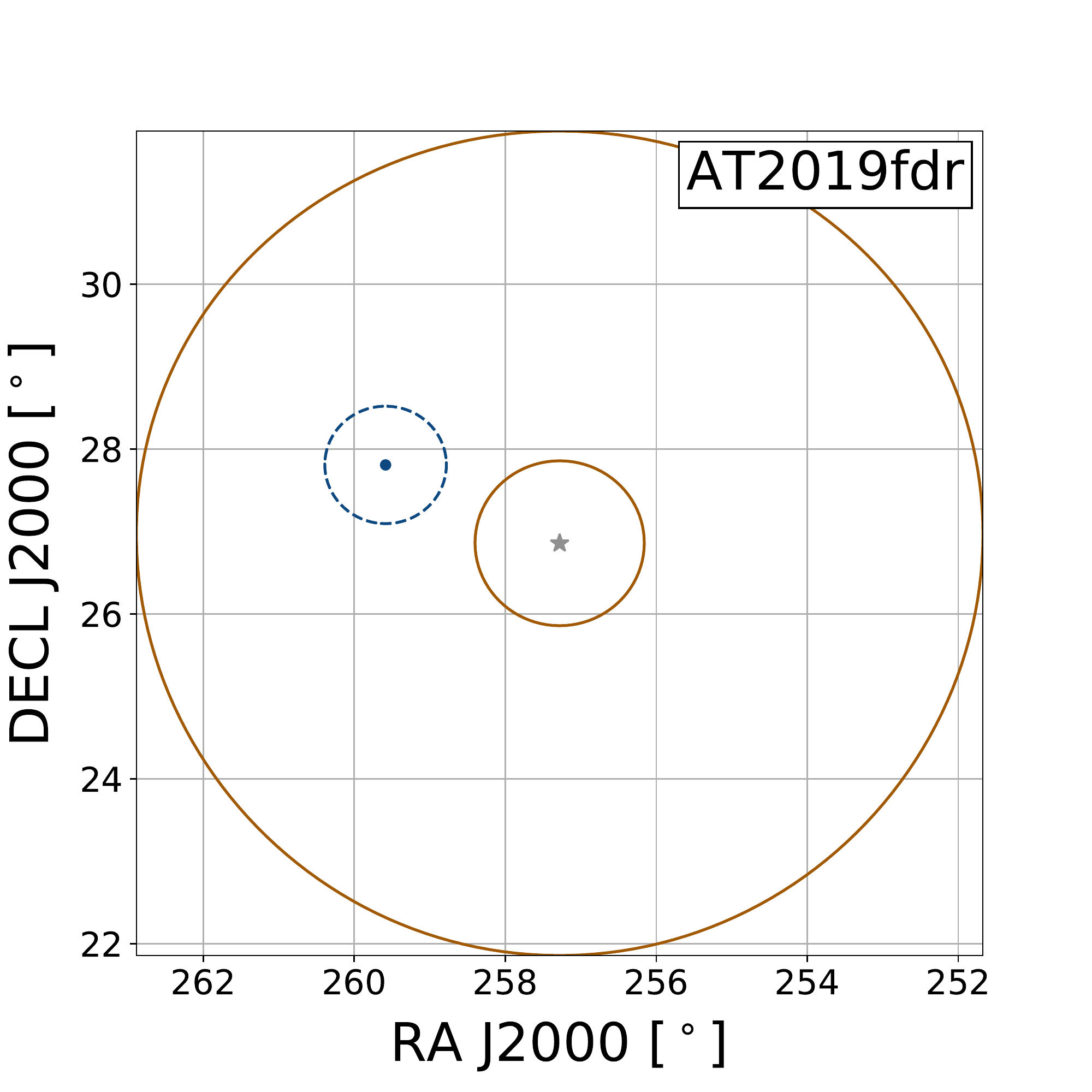}
\put (13,89){\color{black} PRELIMINARY}
\end{overpic}
\caption{Distribution of ANTARES events in equatorial coordinates around the position of AT2019dsg (left) and AT2019fdr (right). The orange lines depict the one and five degree distance from the source position, indicated as a grey star. Only track-like events (blue points) have been detected within $5^{\circ}$ from the location of the sources. The dashed circles around the events indicate the angular error estimate on the track reconstructed direction. 
}
\label{fig:Cluster}
\end{figure*}

\textit{Results - } Figure~\ref{fig:Cluster} shows the distribution of the ANTARES events close to the location of the sources. In both cases, only one event has been detected within $5^{\circ}$ from the TDE. The event close to AT2019dsg (AT2019fdr) is a track-like event, located at a distance of $1.7^{\circ}$ ($2.3^{\circ}$) from the source, and was registered on February~11,~2020 (September~25,~2019). The information on the two events is reported in Table~\ref{tab:events}.

\setlength{\tabcolsep}{.3em}
\begin{table*}[ht!]
        \centering
    \label{tab:events}
    \caption{ANTARES event within an angular distance $\Delta\Psi < 5^{\circ}$ from AT2019dsg (first row) and AT2019fdr (second row). For each ANTARES event, the table reports the topology, equatorial coordinates (RA, $\delta$), angular distance $\Delta\Psi$ from the TDE, Modified Julian Date (MJD), date (dd/mm/year), and fraction $\mathpzc{f}$ of selected data events with energy estimator values larger than that of the event.\medskip}
\resizebox{\linewidth}{!}{
        \begin{minipage}{0.8\textwidth} 
        \centering
 {\def\arraystretch{1.6}
    \footnotesize
    \begin{tabular}{c||c|c|c|c|c|c}
   \hline
      \hline
 Source  & topology & \makecell{(RA, $\delta$) \\ deg} & \makecell{$\Delta\Psi$ \\ deg}  & MJD & \makecell{date \\ dd/mm/year}  & $\mathpzc{f}$ \\
\hline
AT2019dsg & track & (315.8, 13.4) &  1.7 & 58890.99 & 11/02/2020 & 0.35 \\ 
AT2019fdr & track & (259.6, 27.8) &  2.3 & 58751.08 & 25/09/2019 & 0.15 \\ 
\end{tabular}
      }
\end{minipage}
}
\end{table*}

\indent The likelihood method fits less than one signal event at the direction of both TDEs for the three tested values of the spectral index $\gamma = 2.0$, $\gamma = 2.5$, and $\gamma = 3.0$. At the location of AT2019dsg, the largest deviation from the background expectation is obtained for an $E_{\nu}^{-3.0}$ spectrum, with a number of fitted signal events $\hat{\mu}_\mathrm{sig} = 0.7$ and a p-value of $8.9\%$ corresponding to $1.3\sigma$ significance (in the one-sided convention). The lowest p-value ($6.7\%$) at the location of AT2019fdr is observed for an $E_{\nu}^{-2.0}$ spectrum, with a number of fitted signal events $\hat{\mu}_\mathrm{sig} = 0.5$, yielding a significance of $1.5\sigma$ (in the one-sided convention).

Table~\ref{tab:resultsTtot} summarises the results of the searches in terms of best-fit number of signal events and p-value, and reports the upper limits on the one-flavour neutrino flux normalisation, $\Phi_0$, for three values of $\gamma$, where the neutrino flux has been parameterised as $\Phi_{\nu} = \Phi_0 \ \left( \frac{E_\nu}{1\,{\rm GeV}} \right)^{-\gamma}$. Moreover, upper limits are set on the one-flavour neutrino fluence, $\mathcal{F}$, defined as the integral in time and energy of the neutrino energy flux: 

\begin{equation}
    \mathcal{F}  = \int_{t_{\rm min}}^{t_{\rm max}}\int_{E_{\rm min}}^{E_{\rm max}} E_\nu \cdot \Phi_{\nu}\ dE_\nu\ dt = \Delta t \cdot \Phi_0 \cdot \int_{E_{\rm min}}^{E_{\rm max}} E_\nu \cdot \left( \frac{E_\nu}{1\,{\rm GeV}} \right)^{-\gamma}\ dE_\nu,
    \label{eq:fluenceeq}
\end{equation}

\noindent where $E_{\rm min}$ and $E_{\rm max}$ are the boundaries of the energy range containing 90\% of the expected signal events for the given $\gamma$, and $\Delta t$ is the livetime of the search.

The absence of any significant detection of neutrinos from AT2019dsg is consistent with all proposed emission models for the TDE, being their predictions much below the ANTARES neutrino detection threshold~\cite{TDEWinter, TDELiu, TDEMurase}. The expected neutrino fluence resulting from those models is too low to be constrained by the upper limits set in this analysis.

\setlength{\tabcolsep}{.3em}
\begin{table*}[ht!]
        \centering
    \label{tab:resultsTtot}
    \caption{Results of the search at the location of AT2019dsg and AT2019fdr in terms of best-fit number of signal events $\hat{\mu}_\mathrm{sig}$, p-value, 90\%~C.L. sensitivity and upper limits on the one-flavour neutrino flux normalisation $\Phi_{0}^{90\% \mathrm {C.L.}}$ (in units of $\rm{GeV^{-1} cm^{-2} s^{-1}}$), and on the one-flavour neutrino fluence $\mathcal{F}^{90\% \mathrm {C.L.}}$ (in units of $\rm{GeV cm^{-2}}$), for different values of the spectral index $\gamma$. The boundaries, $E_{\rm min}$ and $E_{\rm max}$, of the energy range containing 90\% of the expected signal events, employed in the calculation of the fluence, are listed in the last column.\medskip}
    
    \resizebox{\linewidth}{!}{
        \begin{minipage}{0.9\textwidth} 
        \centering
 {\def\arraystretch{1.6}
    \footnotesize
    \begin{tabular}{c|c||c|c|c|c|c|c|c}
      \hline
            \hline
\multicolumn{2}{c||}{Source} & \multicolumn{7}{c}{Results} \\
   \hline
   
\multicolumn{1}{c|}{Name} & 
\multicolumn{1}{c||}{$\gamma$} & 
\multicolumn{1}{c|}{$\hat{\mu}_\mathrm{sig}$} &
\multicolumn{1}{c|}{p-value} & 
\multicolumn{2}{c|}{\makecell{$\Phi_{0}^{90\% \mathrm {C.L.}}$}} &
\multicolumn{2}{c|}{\makecell{$\mathcal{F}^{90\% \mathrm {C.L.}}$}} &
\multicolumn{1}{c}{\makecell{log$(\frac{E_{\rm min}}{\textrm{GeV}})$ - log$(\frac{E_{\rm max}}{\textrm{GeV}})$ }} \\
    &  &  &  & sensitivity & limit  & sensitivity & limit \\

\hline
\hline
AT2019dsg 
& 2.0 & $<0.1$ & $12.4\%$ & $7.3 \times 10^{-8}$ & $1.0 \times 10^{-7}$ & 14 & 19 & 3.6 - 6.6 \\  
& 2.5 & 0.2 & $10.2\%$ & $1.5 \times 10^{-5}$ & $2.2 \times 10^{-5}$ & 29 & 43 & 2.8 - 5.5 \\ 
& 3.0 & 0.7 & $8.9\%$ & $1.2 \times 10^{-3}$ & $2.0 \times 10^{-3}$ & 230 & 380 & 2.1 - 4.7 \\ 
\hline
AT2019fdr 
& 2.0 & 0.5 & $6.7\%$ & $8.5 \times 10^{-8}$ & $1.3 \times 10^{-7}$ & 15 & 23 & 3.6 - 6.6 \\  
& 2.5 & 0.5 & $7.9\%$ & $2.1 \times 10^{-5}$ & $3.0 \times 10^{-5}$ & 39 & 55 & 2.8 - 5.5 \\ 
& 3.0 & 0.6 & $9.1\%$ & $2.0 \times 10^{-3}$ & $3.0 \times 10^{-3}$ & 360 & 540 & 2.1 - 4.7 \\

\end{tabular}
      }
      \end{minipage}
    }
\end{table*}

\textit{Conclusions - } The ANTARES follow-up of AT2019dsg and AT2019fdr, the TDEs recently indicated as the most likely counterparts of two high-energy IceCube neutrinos, IC191001A and IC200530A, has been presented. The ANTARES data collected since the discovery of AT2019dsg and AT2019fdr have been investigated to look for a steady neutrino emission from the direction of the TDEs, resulting in no evidence for signal in the data. Given the non-detection, 90\%~C.L. upper limits on the one-flavour neutrino flux and fluence have been set. Recent models proposed to explain the multi-messenger observations from AT2019dsg predict a neutrino emission well below the ANTARES sensitivity and can not thus be constrained by the upper limits set in this analysis.
Nevertheless, alternative search methods such as stacking and time-dependent analyses will provide an effective way to enhance the discovery potential using the existing data. Moreover, a significant improvement in sensitivity is expected from the upcoming next generation neutrino telescope, KM3NeT~\cite{LOI}, thanks to its larger effective area combined with an excellent angular resolution~\cite{ARCAPS}.

\vspace{10mm} 

\textit{Acknowledgements -} The authors acknowledge the financial support of the funding agencies:
Centre National de la Recherche Scientifique (CNRS), Commissariat \`a
l'\'ener\-gie atomique et aux \'energies alternatives (CEA),
Commission Europ\'eenne (FEDER fund and Marie Curie Program),
Institut Universitaire de France (IUF), LabEx UnivEarthS (ANR-10-LABX-0023 and ANR-18-IDEX-0001),
R\'egion \^Ile-de-France (DIM-ACAV), R\'egion
Alsace (contrat CPER), R\'egion Provence-Alpes-C\^ote d'Azur,
D\'e\-par\-tement du Var and Ville de La
Seyne-sur-Mer, France;
Bundesministerium f\"ur Bildung und Forschung
(BMBF), Germany; 
Istituto Nazionale di Fisica Nucleare (INFN), Italy;
Nederlandse organisatie voor Wetenschappelijk Onderzoek (NWO), the Netherlands;
Council of the President of the Russian Federation for young
scientists and leading scientific schools supporting grants, Russia;
Executive Unit for Financing Higher Education, Research, Development and Innovation (UEFISCDI), Romania;
Ministerio de Ciencia e Innovaci\'{o}n (MCI) and Agencia Estatal de Investigaci\'{o}n:
Programa Estatal de Generaci\'{o}n de Conocimiento (refs. PGC2018-096663-B-C41, -A-C42, -B-C43, -B-C44) (MCI/FEDER), Severo Ochoa Centre of Excellence and MultiDark Consolider, Junta de Andaluc\'{i}a (ref. SOMM17/6104/UGR and A-FQM-053-UGR18), 
Generalitat Valenciana: Grisol\'{i}a (ref. GRISOLIA/2018/119) and GenT (ref. CIDEGENT/2018/034) programs, Spain; 
Ministry of Higher Education, Scientific Research and Professional Training, Morocco.
We also acknowledge the technical support of Ifremer, AIM and Foselev Marine
for the sea operation and the CC-IN2P3 for the computing facilities.

\bibliographystyle{utphys}
\providecommand{\href}[2]{#2}\begingroup\raggedright\endgroup

\end{document}